\documentclass[fleqn,12pt]{article}

\newcommand{\AmS}{{\protect\the\textfont2
  A\kern-.1667em\lower.5ex\hbox{M}\kern-.125emS}}

\title{Neutrino Geophysics at Baksan (Part II):
Possible Studies of Antineutrino- and Radiogenic Heat Sources in Earth Interior}

\author{G. Domogatsky$^1$, V. Kopeikin$^2$, L. Mikaelyan$^2$, V. Sinev$^2$ \\
\\
$^{1}$Institute for Nuclear Research RAS, Moscow, \\
$^{2}$Russian Research Center "Kurchatov Institute"}
 
\begin{document}

\maketitle


\begin{abstract}
Antineutrinos born inside the Earth (``geoneutrinos'') carry out information of fundamental importance 
for understanding of the origin and evolution of our planet. We show that Baksan Neutrino Observatory 
is one of the best sites for detection and analysis of geoneutrinos using large liquid scintillation spectrometer.
Also we present a short story of concept of Earth as antineutrino source (1960 - 2004 yy).
\end{abstract}

\section*{Introduction}

In this paper we consider future studies at BNO (Baksan Neutrino Observatory of Institute for Nuclear Research RAS) 
of terrestrial antineutrinos $\bar{{\nu}_e}$ coming from beta decay of Uranium and Thorium daughter products 
(``Geoneutrinos''). Geoneutrinos bring information on the Uranium and Thorium abundances and radiogenic heat sources inside 
the Earth, which are of a key importance for understanding of the formation and subsequent evolution of our planet [1].
Geoneutrino is a part of future studies at BNO of low energy $\bar{{\nu}_e}$  fluxes of natural origin aimed for obtaining 
information on their sources, which is otherwise inaccessible. For this purpose a large target mass $\bar{{\nu}_e}$ scintillation 
spectrometer is planned with the inverse beta decay 

\begin{equation}
\bar{{\nu}_e}+p \rightarrow n + e^{+}
\end{equation}
as the detection reaction.
 
 The program also includes:
\begin{itemize}
\item Estimation of frequency of gravitational collapses in the Universe by detection of isotropic flux of $\bar{{\nu}_e}$,  
[2, 3].
\item Test of the hypothesis that a self-sustaining nuclear chain reaction is burning at the center of the Earth (``Georeactor'') 
[4$-$7].
\item Search for solar antineutrinos, which can be produced through $^{8}$B neutrino spin-flavor precession.
\item We note also that no spectroscopic information has so far been obtained on the solar pep, $^{7}$Be and CNO 
neutrinos (${{\nu}_e}$). We hope that in this important field progress can be achieved at BNO with ${{\nu}_e},e$ scattering 
as detection reaction.
\item Quite recently old ideas [8] on using   detection for nuclear reactor control (Plutonium production, nonproliferation etc.) 
have been revived: (See ``Neutrino and Arms Control'' 2004, http://www.phys.hawaii.edu/~jgl/ nacw.html). BNO can join 
international efforts in this field. 
\end{itemize}

KamLAND Collaboration in Japan, using reaction (1) has recently demonstrated revolutionary progress in $\bar{{\nu}_e}$ 
detection technique. KamLAND has already reported that a few geoneutrino events have been observed [9] and moves 
towards detection of soft solar neutrinos. BOREXINO Collaboration at LNGS, Italy, is planning to start soon experiments 
in the same direction. Under the circumstances question can be asked: Is it necessary to start new projects in the same field?

As far as geoneutrinos are concerned, the answer is that their flux is expected to depend strongly on the point of observation. 
The intensity of $\bar{{\nu}_e}$ is predicted to be maximal at high mountains sites (Himalaya, Caucasus where BNO is 
situated), to be minimal in the ocean far from continents (Hawaii, and may be Curacao [7]) and have intermediate values in 
Japan and Italy.

We note also that for success of the experiments listed above the level of the backgrounds is of a decisive importance. To 
a great degree the background is produced by cosmic muons and by $\bar{{\nu}_e}$ flux from power reactors. At BNO 
located at the 4800 mwe rock overburden, the muon flux is much lower than in KamLAND and BOREXINO experiments 
(Fig. 1). The flux of reactor $\bar{{\nu}_e}$ at BNO is also 10 and 2 times lower than it is at mentioned laboratory sites 
respectively [6].

We conclude that BNO is one of the most promising sites to build a massive antineutrino scintillation spectrometer for these 
studies. Essentially we propose to prepare the next step in low energy antineutrino and neutrino physics, which, using 
experience accumulated in KamLAND and BOREXINO projects, will provide further progress in the fields discussed here. 
Importance of this goal is confirmed by recently published idea of LENA experiment [10].

In the next sections we give a short overview of the history of geoneutrino problem, which is now almost 45 years old, next we 
consider geoneutrino models and will finish with the results we hope to obtain at BNO.

\section{Geoneutrinos: First Estimations of Fluxes and\quad Search\quad for\quad Detection\quad Reaction\\ 
(1960-1984)}

Marx and Menyard [11] in 1960 were the first to point out that the Earth is a source of antineutrinos, which are produced in U 
and Th daughter product beta decay, in decay of $^{40}$K and of some other long-lived nuclei. They assumed that natural 
radioactivity is concentrated in a thin upper layer of the Earth and estimated the $\bar{{\nu}_e}$ flux at the surface:

\begin{equation}
F \sim 6.7\times {}10^6 \: {\rm cm^{-2} s^{-1}},
\end{equation}
which is not far from modern values. Radioactivity inside the Earth is unknown. If the concentration were not to decrease with 
the depth, the flux could be hundred times higher.

``In the far future an experiment will be needed to set an upper limit on the antineutrino activity of the Earth. This probably is the 
only way to get information on the composition of the substance in the deep layers of the Earth'' wrote M. A. Markov in 1964 
[12]. Markov, as far as we know, was the first to mention the process (1) as a possible geoneutrino detection reaction. 
``Because of high threshold, $-$ he continued, $-$ the number of active $\bar{{\nu}_e}$ in this case is very low''. He stressed 
also the importance of finding a detection reaction with lower energy threshold. (Typical geoneutrino energy spectrum is shown 
in Fig. 2.)

Two years later Eder [13] explained the gradual slowing down of the circulation rate of the Earth and the increase of its 
radius by a large amount of radiogenic heat sources in the mantel, ``sufficient to blow up our planet''. The relevant geoneutrino 
flux was found two orders of magnitude higher than (2). Eder also mentioned process (1) as a possible detection reaction. No 
details of detector layout and detection principle were discussed at that early time.
	
Antineutrinos of sufficiently low energy can, principally, be detected via resonant reactions of induced orbital electron capture in 
otherwise stable nucleus $A(Z)$ [14]:

\begin{equation}
\bar{{\nu}_e} + e^{-} + A(Z)\rightarrow A(Z-1).
\end{equation}

The reaction signature can be appearance of radioactive nucleus $A(Z-1)$ in the irradiated target (radiochemical method). 
Antineutrinos are captured only in a narrow energy band near the resonant energy $E_{res}\approx T_{max}$, where 
$T_{max}$ is maximal kinetic energy of the electrons in the decay of $A(Z-1)$: 
$A(Z-1)\rightarrow A(Z)+e^{-}+\bar{{\nu}_e}+T$. Clearly, the resonant energy $E_{res}$ is $2mc^{2}=1.02$ MeV 
lower than the threshold energy in the ``usual'' inverse beta decay reaction on the same nucleus:
$$
\bar{{\nu}_e}+A(Z)\rightarrow A(Z-1)+e^{+}.\qquad \qquad \qquad \qquad \qquad \qquad \qquad \eqno (3.1)
$$
The transition probability $w$ per unite time between initial and final states in reaction (3) can be written as:

\begin{equation}
w\sim {(ft)}^{-1}\mid \psi (0)\mid ^{2}dF{(E_{res})}/dE.
\end{equation}

Here  $\mid \psi (0)\mid ^{2}$ is the probability density for atomic electron at nuclear surface, $ft$ is the reduced lifetime of 
beta decay (3.1) and $dF{(E_{res})}/dE$ ${\rm cm^{-2}\,s^{-1}\,MeV^{-1}}$ is the spectral density of incoming 
$\bar{{\nu}_e}$ flux at resonant energy. The density $\mid \psi (0)\mid ^{2}$ rapidly increases with atomic number, 
approximately as $Z^{3}$, the spectral density is high at energies lower than 1.8 MeV (Fig. 2), the inverse beta decay 
reaction (3.1) at high $Z$ is suppressed by positron repulsion in the final state. As a result the resonant capture dominates 
over (3.1) for sufficiently heavy target nuclei.

Potentially, the resonant capture reaction, unlike reaction (1), can search for terrestrial $^{40}$K antineutrinos, which is 
of a vital importance for geophysics.

In 1968 y Markov and Zatsepin organized in Moscow an International Workshop on Neutrino Physics and Astrophysics, 
which later transformed to regular ``NEUTRINO'' Conferences. At this Workshop R. Davis considered for geoneutrino 
detection purposes two resonant reactions:

\begin{equation}
{\rm ^{35}Cl\rightarrow ^{35}\!\!S, {\it E_{res}}= 0.17\: MeV\; and\; ^{209}Bi \rightarrow ^{209}\!\!Pb, {\it E_{res}}=
0.64\: MeV}
\end{equation}
with subsequent radiochemical separation of radioactive $^{35}$S and $^{209}$Pb.

Marx ([15], 1969) considered various models of terrestrial U, Th and K distribution, estimated relevant $\bar{{\nu}_e}$ fluxes 
and recognized the importance of both resonant and inverse beta decay reactions for geoneutrino detection.
 
Krauss, Glashow and Schramm ([16], 1984) calculated radiogenic heat powers, $\bar{{\nu}_e}$ spectra and fluxes due to 
U, Th and K (Table 1).They assumed spherically symmetric distribution of radioactive isotopes in a 100 km thick 
outer layer of the Earth; thus their results correspond to some average point on the surface. Most attention in [16] is given 
to geoneutrino registration via resonant and inverse beta decay reactions; the cross-sections for large number of nuclei have 
been calculated. In the most favorable cases the transition probability (4) was found to be:

\begin{equation}
w\sim (3-6)\times 10^{-31}/{\rm year}.
\end{equation}

\noindent This is several thousand times lower than is observed in solar Ga$\rightarrow$ Ge experiments\dots ``Detection of 
terrestrial antineutrinos will require sophisticated new technology. We can envision several generations of experiments aimed at 
measuring different parts of the antineutrino spectrum''. The first of these generations based on reaction (1) has surpassed in 
2002 y the sensitivity level (6) and started detection of U and Th antineutrinos. Potassium antineutrinos still wait for their 
opportunity.

\section{ Radiogenic\quad Heat -\quad and\quad Antineutrinos\quad Models}
 
Here we widely use information taken from book [1], encyclopedias, from Hofmann papers [17] and Fiorentiny group [18] 
publications. Modern models distribute the Earth's antineutrino- and radiogenic heat sources U, Th and K between 
the crust and the mantle. The Earth's structure schematically shown in Fig. 3 includes components:
\begin{itemize}
\item The crust. Thickness of the continental crust varies from $\sim $70 km (at mountain sites like Hymalaya, Baksan etc.) to 
about 10 km, averaging around $\sim$ 35 km. The continental crust contains a considerable part of the Earth's radioactivity. 
The oceanic crust is 5-6 km thick, covers about $3/4$ of the Earth's surface; the U, Th and K concentrations are 
much lower than in the continental crust. 
\item The outer frontier of the Earth's core lies at the depth of 2900 km. It is believed to be in a molten state because transverse 
seismic waves do not penetrate inside it. Inside the molten core is (partially) crystallized inner core. It is assumed that both 
cores are built up of iron, nickel and some amount of lighter elements; it is believed (by the majority of geochemists) that 
there is no radioactivity in the core, because U, Th and K are chemically ``incompatible'' with the main components.
\item The mantle, which is between the crust and molten core, is subdivided in two layers: the upper and the low. Volcanic 
outflows coming from the upper mantle show, that this layer is strongly ``depleted'' of radioactive elements. The lower mantle 
seems to be not so much depleted, but no systematic information is available on the subject.
\end{itemize}

\bigskip 
How much of U, Th and K is contained in the Earth (crust + mantle)? It is believed that meteorites of a special type 
(carbonaceous chondrites) originate from the same substance that formed the primordial Earth 4.55 billion years ego. 
Starting from this point geochemists deduce the result:

\begin{equation}
a({\rm U}) = 2\times 10^{-8}{\rm g/g},\quad a({\rm Th})/a({\rm U}) = 3.9,
\end{equation}
where $a({\rm U})$ and $a({\rm Th})$ are element mass abundances.

Using (7) one can find total mass of Uranium $M$(U) and total mass of Thorium $M$(Th), contained in the crust + mantle
 and can calculate relevant radiogenic heat powers $H$(U) and $H$(Th):

$$
{\rm Crust + Mantle: {\it M}(U) = 0.8\times10^{17} kg, \,\, {\it M}(Th) = 3.15\times 10^{17}kg},
$$
 
\begin{equation}
{\rm Crust + Mantle: {\it H}(U) = 7.6\, TW,\, {\it H}(Th) = 8.3\, TW} 		
\end{equation}
 
Some features of U and Th decay are given in Tables 2, 3. 
The Earth in this model radiates $L_\nu = 1.1\times 10^{25}\, {\bar{{\nu}_e}}/s$ emitted in U- and Th decay chains, 
which carry away $1.1$ TW. 

The outlined model (sometimes called Bulk Silicate Earth, BSE) is shared by a great majority of geochemists and geophysicists.

However, the BSE prescription does not work with K. Average K to U concentration ratio in carbonaceous chondrites 
is $\sim$ 6 higher than is typical for terrestrial samples. This problem has been debated for decades. Two ideas have been 
proposed to explain K deficit in Earth's crust + mantle.
\begin{enumerate}
\item  Tiny particles of which the Earth was formed 4.55 billion years ago had at that time higher temperatures than chondrites 
and had lost most part of original K through evaporation. In this case K contribution to the radiogenic heat is 
{\it H}(K) $\sim$ 2.5 TW.
\item  Actually there is no K deficit at all; just the major part of primordial K has gone into the core. In this case 
{\it H}(K) $\sim$ 16 TW, most part of this power comes from the core. Recent laboratory experiments provide evidence that 
actually large quantity of K could enter the iron rich Earth's core [19].
\end{enumerate}

Questioned is the theory that there is no U in the Earth's core. As already mentioned a self-sustaining fission reaction is 
supposed to burn there with a power release of 3$-$9 TW [4]. Also questionable is the type of meteorites, which ought to be 
used as a starting point for finding the amount of Earth's radioactivity [20].

In the BSE model the total radiogenic power $H_{rad} = H({\rm U}) + H({\rm Th}) + H({\rm K})\approx 19$ TW is 
slightly lower than one half of the total Earth heat outflow ($H_{tot}\approx 40-42$ TW). As an upper limit sometimes is 
considered a ``fully radiogenic'' model with U etc. masses increased by a factor $\sim$ 2.2 so that $H_{rad} = H_{tot}$.

\section{Strategy of Research, Geoneutrino Fluxes, Observables}

Progress in the low energy $\bar{{\nu}_e}$ detection technique (the CHOOZ experiment, KamLAND and BOREXINO 
proposals) opened practical opportunities to detect geoneutrinos. In 1997-1998 yy Rothschild, Chen and Calaprice [21] 
and Raghavan et al. [22] proposed to use large target mass liquid scintillation detectors to investigate the geoneutrinos with 
the inverse beta decay (1) as a detection reaction.

Authors of [21] recognized that measurements are to be done at a number of geographical positions:
\begin {itemize}
\item in mountain regions where contribution of core radioactivity is dominant and geoneutrino intensity is expected to be 
maximal, 
\item in the ocean far from continents with dominant contribution from the mantle and minimal expected geoneutrion flux and 
\item at Kamioka site in Japan and at LNGS in Italy with an intermediate level of geoneutrino intensity. 
\end {itemize}

Since 2002 y Fiorentini group in Italy has started extensive calculations of expected geoneutrino fluxes, detection rates and 
their dependence on the point of observation on the Earth's surface (see e.g. [18]).

In scintillation detectors geoneutrino energy spectrum at $E >1.806\,$ MeV is measured using delayed coincidences between 
spatially correlated positron and neutron produced in reaction (1). $T\approx E-1.806$ gives relation between positron 
kinetic energy $T$ and the energy of incoming $\bar{{\nu}_e}$. In large target mass detectors the positron annihilation quanta 
are absorbed in the feducial volume and positron energy release $E_{vis}$ is :

\begin{equation}
E_{vis}\approx E - 0.8\, \, ({\rm MeV}),
\end{equation}

Thus the minimal positron energy release at the reaction threshold when $T$ = 0 is $E_{vis(min)} = 1.02$ MeV.

\bigskip 
The numbers of Uranium $N$(U) and Thorium $N$(Th) geoneutrino events produced in the scintillator target exposed 
to incident $\bar{{\nu}_e}$ fluxes $F$(U) and $F$(Th) are given by:

$$
N({\rm U) /year\, 10^{32} H} = 13.2\, F({\rm U)\times 10^{-6}\, cm^{-2} s^{-1}}\qquad \qquad \qquad \qquad   
$$
 
\begin{equation}
N({\rm Th)/year\, 10^{32} H} = 4.1\, F({\rm Th)\times 10^{-6}\, cm^{-2}s^{-1}}		
\end{equation}
(1 160 ton of ${\rm CH_{2}}$ – based scintillator contains $10^{32}\,$H atoms)

Geoneutrino fluxes $F$(U) from Uranium decay calculated in BSE model in [18] and [21] for a number of sites are shown 
in Fig. 4. In [18] higher values of Uranium abundances in the continental crust are assumed than those used in [21]. As a result 
differences in calculated fluxes for the same point of observation are rather large, which demonstrates good sensitivity of the 
neutrino method to the assumption on the distribution of radioactivity in the Earth. In Fig. 4 one can also clearly see a rapid 
decrease of the flux when the point of observation shifts from the mountains to the oceanic site.

The total numbers of the Uranium and Thorium events $N$(U) + $N$(Th) (Table 4) also demonstrate large scattering of 
predictions found in [18,21,22]. The expected number of geoneutriono events at BNO near the Elbrus mountain is close 
to that found for geographical maximum in Himalayas. As can be seen in Fig. 5 Uranium and Thorium events can be separated 
and thus the ratio of U/Th abundances measured. One also can see that at BNO the background coming from nuclear 
power reactors is sufficiently low.

The size and construction of the future detector and backgrounds require special consideration. Partially these issues have been 
considered in our recent publication on georeactor [6]. Good passive shield should protect the scintillator from the natural 
radioactivity of the surrounding rock, from PhM's glass and metallic support constructions. The anticoincidence system should 
effectively tag cosmic muons and showers. As already mentioned in the Introduction the muon flux at BNO is considerably 
lower than at Kamioka and LNGS (Fig. 1). Experience accumulated in KamLAND and BOREXINO projects shows that 
deepest purification of liquid scintillator can be achieved and extremely low level of radioactive contaminations can be reached, 
such as $\sim 3\times 10^{-18}$ U$\;$g/g.

In [6] we have considered LS volume sufficiently large $(10^{32}$ H atoms,\quad 1.16 kton) to confirm or reject theory that 
a nuclear reactor is burning at the center of the Earth. The geoneutrino studies seem to require larger target mass detectors, 
may be as large as $(1-2)\times 10^{33}$ H atoms. Indeed, (a) the detection rates due to oscillation effect will be $\sim$ two 
times lower than those given in the Table 4; (b) good statistics is needed to separate U- and Th geoneutrinos (Fig. 5) 
and, finally, (c) we consider here a new generation experiment, which will operate after KamLAND. With an exposition 
$(50-100)\times 10^{32}\,{\rm H\cdot year}$ it would be possible to accumulate at BNO 2000-4000 geoneutrino events. 
With statistics like this one can try to investigate the angular distribution of incoming geoneutrinos, which can give valuable 
information on the location of U and Th masses in the interior of the Earth.

\section* {Conclusion}

Uranium and Thorium distribution in the Earh's interior and relevant radiogenic heat can be found via geoneutrino observations 
at a number of geographical points.
 
The Baksan Neutrino Observatory is one of the most promising sites to build a massive antineutrino scintillation spectrometer 
for geoneutrino studies and also for investigations on low energy $\bar{{\nu}_e}$ and ${{\nu}_e}$ fluxes of natural origin 
aimed for obtaining information on their sources, which are otherwise inaccessible.
 
In 1984 y Krauss, Glashow, and Schramm wrote: ``'Neutrino and antineutrino astronomy and geophysics can open vast new 
windows for exploration above us and below''. The first window is now opening at Kamioka. New windows can be opened 
at Baksan and elsewhere. Detector for Potassium antineutrinos is still to be invented.

\section*{Acknowledgments}
This work is supported by RF President grant 1246.2003.2 and by RFBR grant 03-02-16055. Authors thank Profs. 
G. Fiorentini and E. Lisi for useful comments.

\vspace{6.5 cm}

\begin{table}[htb]
\caption{Potassium, Thorium and Uranium in the Earth's lithosphere: Masses ($M$), fluxes of $\bar{{\nu}_e}$ ($F$) and 
radiogenic heat powers ($H$) according to$\;$[16]}
\label{table}
\vspace{5pt}
\hspace{70pt}
\begin{tabular}{|c|ccc|} 
\hline
Element & $M$ & $F$ & $H$ \\
 & $10^{17}$ kg & $10^{6}\,{\rm cm^{-2} s^{-1}}$& $10^{12}$ W \\
\hline
U & 0.80 & 3.5 & 9.6 \\
Th & 3.8 & 3.5 & 7.3 \\
K & $5.2\times 10^{3}$ & 11 & 1.8 \\
\hline
& Total: & 18 & 19 \\
\hline
\end{tabular}
\end{table}

\begin{table}[htb]
\caption{Energy balance (MeV/decay) in equilibrium $^{238}$U and $^{232}$Th decay chains:
Total energy $E_{tot}$, thermal (radiogenic) energy $E_{H}$ and energy carried away by antineutrinos $E_{\nu}$}
\label{table}
\vspace{5pt}
\hspace{55pt}
\begin{tabular}{|c|ccc|}
\hline
& $E_{tot}$ & $E_{H}$ & $E_{\nu}$ \\
\hline
$^{238}$U$\,\Longrightarrow\, ^{206}$Pb$\,+\,8\, ^{4}$He$\,+\,6\,\bar{{\nu}_e}$ & 51.7 & 47.7 & 4.0 \\
$^{232}$Th$\,\Longrightarrow\, ^{208}$Pb\,+\,$6\, ^{4}$He\,+\,4\,$\bar{{\nu}_e}$ & 42.7 & 40.0 & 2.3 \\
\hline
\end{tabular}
\end{table}

\begin{table}[htb]
\caption{Specific radiogenic heat powers $H$, neutrino luminosities $L_{\nu}$ and radiated antineutrino powers $W_{\nu}$ }
\label{table}
\vspace{5pt}
\hspace{30pt}
\begin{tabular}{|c|ccc|}
\hline
& $H, {\rm TW}/10^{17}$kg & $L_{\nu},10^{24}{\rm s^{-1}}/10^{17}$kg & $W_{\nu},{\rm TW}/10^{17}$kg \\
\hline
$^{238}$U & 9.5 & 7.6 & 0.81 \\
$^{232}$Th & 2.6 & 1.6 & 0.15 \\
\hline
\end{tabular}
\end{table}

\begin{table}[htb]
\caption{Expected rates of U + Th events at various sites (exposition $10^{32}\;{\rm H\cdot year}$, no oscillations)}
\label{table}
\vspace{5pt}
\hspace{20pt}
\begin{tabular}{|c|ccc|}
\hline
Site & Rothschild {\it et al.$^{*}$} & Raghavan {\it et al.} & Fiorentini {\it et al.}\\
& [21] & [22] & [18]\\
\hline
Himalaya & 65 & - & 112 \\
Baksan & - & - & 91 \\
Gran Sasso & 53 & Ia: 134; Ib: 50 & 71 \\
Kamioka & 48 & Ia: 75; Ib: 50 & 61 \\
Hawaii & 27 & - & 22 \\
\hline
\end{tabular}\\[2pt]
{\small $^{*}$ Calculated using Eqs. (10) and $\bar{{\nu}_e}$ Fluxes from Ref. [21]}
\end{table}

\end{document}